\documentstyle[epsfig,twocolumn,prl,aps]{revtex}

\begin{document}
\draft

\twocolumn[\hsize\textwidth\columnwidth\hsize\csname @twocolumnfalse\endcsname

\title{Nuclear spin relaxation rates in two-leg spin ladders}
\author{F. Naef and Xiaoqun Wang}

\address{
Institut Romand de Recherche Num\'erique en Physique des
Mat\'eriaux (IRRMA), \\
INR-Ecublens, CH-1015 Lausanne, Switzerland}

\maketitle
\begin{abstract}
Using the transfer-matrix DMRG method, we study the nuclear spin 
relaxation rate $1/T_1$ in the two-leg s=$1/2$ ladder as function 
of the inter-chain ($J_{\perp}$) and intra-chain ($J_{\|}$) couplings.
In particular, we separate the $q_y=0$ and $\pi$ contributions
and show that the later contribute significantly to the copper
relaxation rate $^{63}(1/T_1)$ in the experimentally relevant 
coupling and temperature range. We compare our results to both
theoretical predictions and experimental measures on ladder materials.

\end{abstract}
]

Dynamical properties, while providing the most detailed information on the
physics of low-dimensional antiferromagnets
are also the most difficult for theoretical predictions.
Measured in NMR experiments probing the low-frequency spin dynamics, 
they continue to reveal unexpected features.
For instance, the recent experiments by Imai {\it et al.}\cite{imai} on the 
${\rm Cu}_2{\rm O}_3$ two-leg s$=1/2$ ladders in 
${\rm La}_6{\rm Ca}_8{\rm Cu}_{24}{\rm O}_{41}$
have shown the different temperature $T$ dependence of the 
nuclear spin relaxation rate $1/T_1$ for the oxygen ($^{17}(1/T_1)$) 
and copper ($^{63}(1/T_1)$) atoms. Although both exhibit activated behavior below
$T\simeq 350~{\rm K}$, $^{63}(1/T_1)$ shows a crossover to a linear $T$
dependence above $T\simeq 350~{\rm K}$. Shortly after these experiments were
performed, Ivanov and Lee\cite{ivanov} proposed that the 
crossover may be understood from the momentum transfer
${\bf q}=(\pi,\pi)$ contributions in the weakly coupled chain limit.
The importance of the $q_y=\pi$ processes was also mentioned earlier by
Sandvik {\it et al.}\cite{sandvik} in a study of $^{63}(1/T_1)$ for SrCu$_2$O$_3$,
where they consider the isotropic ladder at $T/J\geq 0.2$.
Despite these findings, a better theoretical understanding of $1/T_1$
in isolated ladders is necessary, as emphasized in Refs.\cite{imai,ivanov}.

Our purpose is to determine $1/T_1$ for the experimentally relevant
ratio of inter ($J_{\perp}$) and intra-chain ($J_{\|}$) couplings,
separating the $q_y=0$ and $\pi$ contributions. We will use the
transfer-matrix density-matrix renormalization group
for the evaluation of thermodynamic properties\cite{wang1} and local
imaginary time correlations\cite{japs,nwzvl} of one-dimensional tight-binding
models. Combined with an analytical continuation, this technique has recently
proven to provide reliable real frequency correlations down to low 
temperatures\cite{nwzvl}. In the low-$T$ regime, our results show a good
agreement to analytical results obtained in the weak or strong coupling
limits\cite{ivanov,troyer}. At higher $T$, we demonstrate the importance of the
$(1/T_1)_{\pi}$ contributions to $^{63}(1/T_1)$ experiments, in particular,
our result reproduces the crossover to the linear paramagnetic regime observed
in the $^{63}(1/T_1)$ rate.

The most recent estimates of $J_{\|}/J_{\perp}$ from susceptibility and
neutron scattering experiments agree for a value of about $0.5$
\cite{johnston,eccleston}. 
On the other hand, estimating $J_{\|}/J_{\perp}$ from NMR rates, we found
that the standard ladder, together with the hyperfine couplings 
measured in Refs.\cite{imai,ishida}, are unable to account for
$J_{\|}/J_{\perp}\sim 0.5$ but favors $J_{\|}/J_{\perp}\sim 1$.

The two-leg ladder is described
by the antiferromagnetic Heisenberg Hamiltonian
\begin{eqnarray}
H\hspace{-0.1cm}=\sum_{i=1}^N J_{\|}({\bf S}_{1,i}\cdot {\bf S}_{1,i+1}
\hspace{-0.1cm}+{\bf S}_{2,i}\cdot {\bf S}_{2,i+1})
\hspace{-0.1cm}+J_{\perp} {\bf S}_{1,i}\cdot {\bf S}_{2,i},
\label{hamiltonian}
\end{eqnarray}
where ${\bf S}_{n,i}$ denotes a s=$1/2$ spin operator at the
$i$-th rung and the $n$-th chain.
It is by now well established that the spectrum of the
Hamiltonian (\ref{hamiltonian})
consists of an $S=0$ ground state, the lowest-lying excited states forming a
gaped $S=1$, $k_y=\pi$ single magnon branch with minimum $\Delta$ at
$k_x=\pi$\cite{troyer,barnes,tsvelik}.

\begin{figure} 
\epsfxsize=2.5 in\centerline{\epsffile{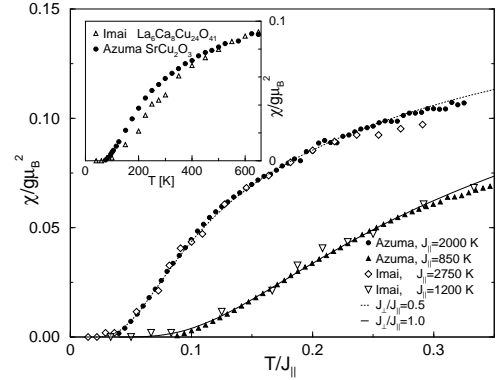}}
\caption{Comparison of the susceptibility and Knight shifts
to transfer-matrix DMRG results for $J_{\|}/J_{\perp}=0.5$
and $1$. Inset: susceptibility and Knight shift measurements compared.}
\label{fig0}
\end{figure}

The issue of determining the values of $J_{\|}$ and $J_{\perp}$
has raised some controversy, indeed estimates from various
authors\cite{imai,azuma,ishida,johnston,mueller} range from 
$J_{\perp}/J_{\|}\simeq 0.5$ to $1$, with an emerging consensus
for about $0.5$. Fitting our numerical results to susceptibility\cite{azuma}
and Knight shift\cite{imai} experiments, we confirm in Fig. \ref{fig0} that the best
agreement is found when $J_{\perp}/J_{\|}=0.5$ for SrCu$_2$O$_3$\cite{johnston_p}, 
implying energy scales $\Delta\simeq 420$K and $J_{\|}\simeq 2000$K.
Nevertheless, a reasonable fit to $J_{\perp}/J_{\|}=1$ is also possible in a slightly 
narrower $T-$range, showing that the susceptibility is relatively
insensitive to $J_{\perp}/J_{\|}$ at low$-T$. Moreover, we point out in
the inset that the susceptibility and Knight shift measurements are not
fully consistent through the entire $T-$range, implying material-dependent
$J_{\|}$ values.  In principle, Knight shifts are
better suited for a comparison with theoretical results, since 
they involve no subtraction of a Curie term and no unknown g
factor\cite{azuma,johnston}.

The nuclear relaxation rate $1/T_1$ is given by
\begin{eqnarray}
&&\begin{array}{l}
{\small b}\\
~
\end{array}
\hspace{-0.2cm}\left( \frac 1 T_1\right)=\sum_{q_y=0,\pi} \int~^bF({\bf q})
 S({\bf q},\omega_N)~dq_x,~~b=63,17\\
&&S({\bf q},\omega)=\frac 1 {2N}\sum_{i,j,m,n}
\int \langle S^z_{n,i}(t)S^z_{m,j}\rangle
e^{i ({\bf q}\cdot (j-i,m-n)+\omega t)}dt~.\nonumber
\label{T1def}
\end{eqnarray}
Here, $\omega_N$ is the nuclear Larmor frequency and $^bF({\bf q})$ are
the appropriate hyperfine couplings. According to Ref.\cite{imai},
$^{63}F({\bf q})=A^2$ for $^{63}{\rm Cu}$,
and $^{17}F_2({\bf q})=4~F^2 \cos^2(q_y/2)$ for the
rung ${\rm O}(2)$ oxygen atoms\cite{defhyp}.
Contrary to CuO$_2$ planes, it has been argued that considering only a
local hyperfine interaction for the $^{63}{\rm Cu}$ nucleus in
ladders\cite{sandvik,ishida} is sufficient.

To determine $1/T_1$, the transfer-matrix DMRG method permits
a very precise evaluation of the imaginary time
Green's function\cite{nwzvl}
\begin{equation}
G(\tau)=\langle S^z_{1,1}(\tau) S^z_{m,j}\rangle
\end{equation}
where $S^z_{n,i}(\tau)=e^{\tau H} S^z_{n,i} e^{-\tau H}$ and the indices
$(m,j)\in \{(1,1),(2,1),(1,2),(2,2)\}$ run over the four corners
of a plaquette. For all calculations,
we chose an imaginary time slice $\epsilon=0.025/J_{\|}$ and $m=200$ states
were kept to represent the transfer-matrix. After an analytical 
continuation using the Maximum Entropy method, we can resolve $S({\bf q},\omega)$
in $q_y=0$ or $\pi$ and obtain the averages over the
$q_x$ momentum transfer:
\begin{eqnarray}
&&{\bar S}_0(q_y,\omega)=\frac 1 {4\pi}\int dq_x 
\cos^2(q_x/2) S(q_x,q_y,\omega)\nonumber\\
&&{\bar S}_\pi(q_y,\omega)=\frac 1 {4\pi}\int dq_x 
\sin^2(q_x/2) S(q_x,q_y,\omega)~.
\end{eqnarray}
As indicated by the notation, ${\bar S}_0(q_y,\omega)$ (${\bar S}_\pi(q_y,\omega)$)
is dominated by processes with $q_x$ close to $0$ ($\pi$).
For the hyperfine couplings given above,
the copper and oxygen rates are fully determined by appropriate
combinations of ${\bar S}_{{\bar q}_x}(q_y,\omega)$, ${\bar q}_x=0,\pi$.
Accordingly, we also define 
\begin{equation}
\left(1\over  T_1\right)_{q_y}={\bar S}_0(q_y,\omega_N)+{\bar S}_\pi(q_y,\omega_N)~,
\end{equation}
the dimensionless contributions to $1/T_1$ from the $q_y=0,\pi$ momentum space 
sectors.

\begin{figure}
\epsfxsize=2.6 in\centerline{\epsffile{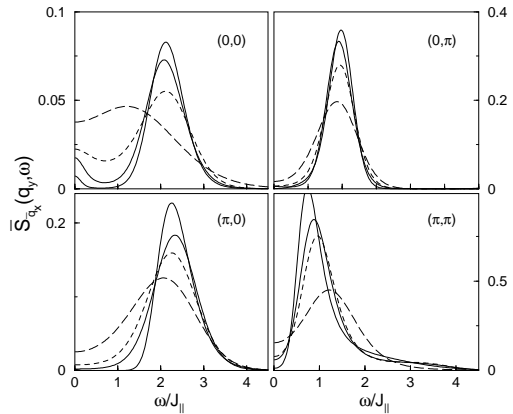}}
\caption{${\bar S}_{{\bar q}_x}(q_y,\omega)$ for $J_{\perp}/J_{\|}$$=$$1$. 
$T/J_{\|}$$=1/2$ (dotted line), $1/3$ (dashed), $1/4$ (thin) and $1/6$
(thick).}
\label{fig1}
\end{figure}

For illustration, we show
in Fig. \ref{fig1} ${\bar S}_{{\bar q}_x}(q_y,\omega)$ for the case $J_{\perp}/
J_{\|}$$=$$1$ as a function of $T$. As the latter
drops, we can see clearly the signatures of the low-lying spectrum
emerging. Indeed, the main peak is due to excitations from the ground 
state to the single magnon branch for the $q_y$$=$$\pi$ sector and to the 
$2-$magnon continuum for $q_y$$=$$0$, for instance, it appears correctly
from ${\bar S}_\pi(\pi,\omega)$ that the minimum gap approaches
the $T$$=$$0$ value $\Delta$$=$$0.502$.
Such contributions to ${\bar S}_{{\bar q}_x}(q_y,\omega)$
become $T$ independent for $T$$\rightarrow$$ 0$,
however, they do not contribute to $1/T_1$ as they all involve
frequencies $\omega$$\ge$$\Delta$$\gg$$\omega_N~(\approx 3~{\rm mK})$.
On the contrary, there are thermally activated contributions
to $1/T_1$, as they involve excitations for $\omega$$\rightarrow$$ 0$\cite{omega}.
As seen in Fig. \ref{fig1}, this limit is dominated in the low-$T$ regime
by processes with $({\bar q}_x,q_y)$$=$$(0,0)$ and $(\pi,\pi)$.

The first contributions involve two thermally excited magnons near the 
minimum of the branch $e(k_x,k_y$$=$$\pi)$$=$$\Delta$$+$$a(k_x-\pi)^2$, 
according to Ref.\cite{troyer}, they lead to
\begin{equation}
\left(\frac 1 T_1 \right)_0~\propto~\frac 1 a e^{-\Delta/T}~
[0.809 -\ln(\frac {\omega_N} T)]~.
\label{T1troyer}
\end{equation}
Corrections to the quadratic minimum\cite{melzi} only change
the prefactor of the exponential term slightly for $T/\Delta\lesssim 0.3$.
The logarithmic divergence in Eq.(\ref{T1troyer}) cannot be resolved by the Maximum 
Entropy method, however, it does not change the dominant exponential
behavior. For instance, the factor in the square
brackets changes only by about $5$ percent from $T=\Delta/2$ to $\Delta/4$.
 
The second important contributions with $q_y=\pi$
involve scattering of single magnons with
the $2-$magnon continuum. By representing the low-lying excitations
in terms of free massive fermions\cite{tsvelik} in the weakly coupled
chain limit, it was proposed in Ref.\cite{ivanov} that
\begin{equation}
\left({1\over T_1}\right)_\pi\propto~
\left({T\over\Delta}\right) e^{-{2\Delta/T}}~.
\label{T1lee}
\end{equation}
As the scale $2\Delta$ corresponds to the bottom of the $2$-magnon continuum at
${\bf k}=(0,0)$ independent of $J_{\perp}/J_{\|}$\cite{barnes,wang_frus}, this 
result should hold over a wider coupling range.
In fact, exact diagonalisation (ED) results show the existence
of large matrix elements $|\langle n|S^z_q|m\rangle|^2$ between 
the $2-$magnon continuum at ${\bf k}=(0,0)$ and the single magnon
branch. Such processes are characterized by a momentum transfer $q_x$ 
which rapidly shifts close to $\pi$ with decreasing $J_{\perp}$.
From the analysis of the spectrum\cite{barnes}, it follows that 
$q_x/\pi\approx 0.5,0.8$ and $0.95$ for $J_{\perp}/J_{\|}$$=$$2,1$
and $0.5$. Other important $q_y=\pi$ processes occur
between the single magnon branch at ${\bf k}$$=$$(0,\pi)$ and the continuum near
${\bf k}$$=$$(\pi,0)$ when $J_{\perp}/J_{\|}\lesssim 0.5$. However, these have 
a larger activation gap of about $3.7\Delta$.
 
\begin{figure}
\epsfxsize=2.8 in\centerline{\epsffile{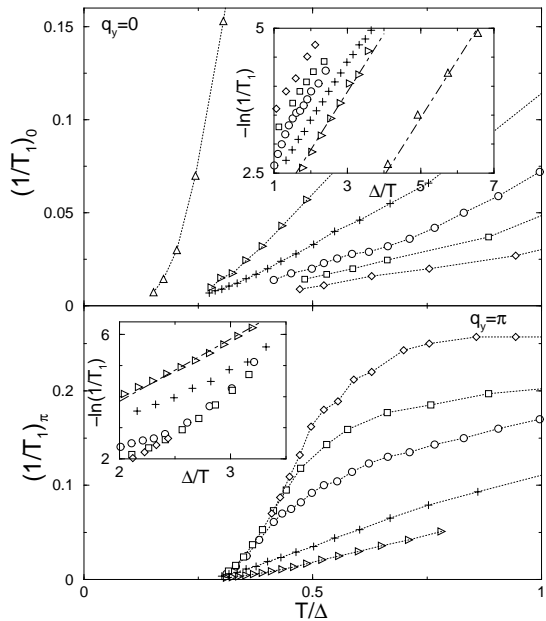}}
\caption{Low-$T$ behavior of $(1/T_1)_0$ and $(1/T_1)_\pi$
for $J_{\perp}/J_{\|}=5$ ($\triangle$), $2$ ($\rhd$), $1.5$ (+),
$1$ ($\circ$), $0.8$ ($\Box$) and $0.6$ ($\Diamond$). Insets:
The dot-dashed lines have slopes $1$ ($2$) for $q_y$$=$$0$ ($\pi$).}
\label{fig2_lt}
\end{figure}

Let us first discuss our results for $1/T_1$ as a function of
$J_{\perp}$ and $T$. We should point out that the values
${\bar S}_{{\bar q}_x}(q_y,\omega\rightarrow 0)$ 
become very small compared to the magnitude of the main peak as $T$
is lowered. Therefore, for a reliable estimate of $1/T_1$, it is essential,
(i) to work with very precise imaginary time
data, and (ii) to separate the different ${\bf q}$ space contributions,
as the integrated weight of the $q_y=0$ and $\pi$ sectors differ considerably.
In Fig. \ref{fig2_lt}, we present $(1/T_1)_0$
and $(1/T_1)_\pi$ in the low-T regime, where $T$ is scaled 
by the spin gap $\Delta$.

$(1/T_1)_0$ shows the correct behavior of Eq.(\ref{T1troyer}),
especially for large $J_{\perp}/J_{\|}$, when low enough $T/\Delta$
can be reached to clearly identify the activated regime.
This is verified in the inset where we have
plotted a line of slope one on a logarithmic scale. 
We also observe that the prefactor of the exponential strongly depends 
on $J_{\perp}/J_{\|}$, consistently with the factor $1/a$ in 
Eq.(\ref{T1troyer}). Indeed, as the dispersion of the single magnon branch
flattens with increasing $J_{\perp}$, $1/a$ grows and 
$\lim_{J_{\perp}\rightarrow\infty}1/a=\infty $\cite{troyer,barnes}.
The $q_y=\pi$ results reveal a larger gap which can be fitted consistently
to Eq.(\ref{T1lee}) when $J_{\perp}/J_{\|}=1.5,2$ (see inset). When
$J_{\perp}/J_{\|}\lesssim 1$, we find some deviations from Eq.(\ref{T1lee})
In this regime, the low-$T$ extraction of $(1/T_1)_\pi$ becomes more
delicate as the growing main frequency peak near $\omega=\Delta$ (due to excitations
from the ground state to the single magnon branch) may bias the 
limit $S({\bf q},\omega\rightarrow 0)$ (see Fig. \ref{fig1}). On the other hand,
we cannot exclude that the above mentioned processes with a gap $\approx 3.7\Delta$
contribute significantly at our lowest temperatures.
For the case $J_{\perp}/J_{\|}=5$ (not shown in Fig. \ref{fig2_lt}), $(1/T_1)_\pi=0$ since
the single magnon band and the $2-$magnon continuum do not yet overlap. 
\begin{figure}
\epsfxsize=2.8 in\centerline{\epsffile{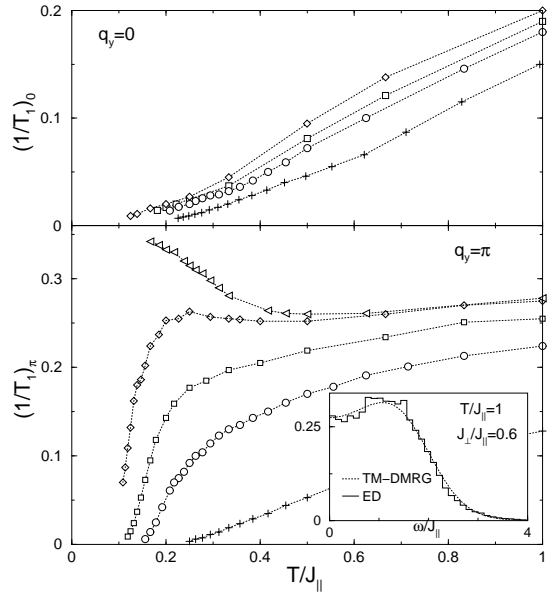}}
\caption{
$(1/T_1)_0$ and $(1/T_1)_\pi$ as function of $T/J_{\|}$
for $J_{\perp}/J_{\|}=1.5$ (+), $1$ ($\circ$), $0.8$ ($\Box$),
$0.6$ ($\Diamond$) and $0.4$ ($\lhd$).
Inset: comparison of ${\bar S}(0,\pi,\omega)+{\bar S}(\pi,\pi,\omega)$
from ED and Transfer-Matrix DMRG.}
\label{fig3_ht}
\end{figure}

A wider $T$ range is shown in Fig. \ref{fig3_ht}.
$(1/T_1)_\pi$ obtained at $T/J_{\|}$$=1$ agree within 5 percent with those
from ED of an $8$-rung ladder, as seen in the inset for $J_{\perp}/J_{\|}$$=0.6$.
However, due to finite size effects, the determination of $1/T_1$ from ED becomes
meaningless below $T/J_{\|}\simeq 0.8$.

As $(1/T_1)_0$ obeys a predominantly linear intermediate to high-$T$ behavior for 
all $J_{\perp}$ above $T\simeq 0.4~J_{\|}$,
$(1/T_1)_\pi$ exhibits a qualitative change as $J_{\perp}/J_{\|}$ is decreased.
Indeed, for $J_{\perp}/J_{\|}\leq 0.6$ $(1/T_1)_\pi$ develops a maximum
near $T/J_{\|}$$\sim 0.2$ becoming sharper with decreasing
inter-chain coupling. We believe such a feature is reminiscent of the weak
coupling limit, as our $(1/T_1)_\pi$ for $J_{\perp}/J_{\|}\leq 0.6$ shows
a good agreement to the behavior predicted in Ref. \cite{ivanov}.
There, Ivanov and Lee also argued that with increasing $T$, $(1/T_1)_\pi$
flattens to the single Heisenberg chain result\cite{sachdev}, as we observe
for $J_{\perp}/J_{\|}=0.4$ and $0.6$ above $T/J_{\|}=0.4$.
In spite of the larger gap, we find that $(1/T_1)_\pi$ dominates
over $(1/T_1)_0$ in a wide intermediate-$T$ regime as can be 
verified from Figs. \ref{fig2_lt} and \ref{fig3_ht}. For
instance, when $T/\Delta=1$, $(1/T_1)_{\pi}$ exceeds
$(1/T_1)_0$ by a factor $2.5$ (resp. $9$) for $J_{\perp}/J_{\|}=1$ (resp. $0.6$).
In particular, for $J_{\perp}/J_{\|}=1$, we observe the
crossing of $(1/T_1)_{\pi}$ and $(1/T_1)_0$ at $T/\Delta\simeq 0.35$.
This consideration implies  that both $q_y$$=0$ and $\pi$ contributions 
are relevant to copper $^{63}(1/T_1)$ experiments above $T\simeq0.35~\Delta$
(typically $\approx 200~{\rm K}$ in Cu$_2$O$_3$ ladders).
In fact, taking into account the usually omitted $q_y$$=\pi$ processes
to fit the low-$T$ $^{63}(1/T_1)$ data of Ishida {\it et al.}\cite{ishida}
on ${\rm SrCu}_2{\rm O}_3$, we obtain for $T$ between $100$ and $300~{\rm K}$
a gap $\Delta\approx 520~{\rm K}$, while considering only the form for $(1/T_1)_0$
leads to $\Delta\approx 700~{\rm K}$. Hence, the discrepancy between estimates
of the gap from $^{63}(1/T_1)$ and susceptibility\cite{azuma,ishida}
(leading to $\Delta\simeq 420$K) is reduced.

\begin{figure}
\epsfxsize=3.2 in 
\centerline{\epsffile{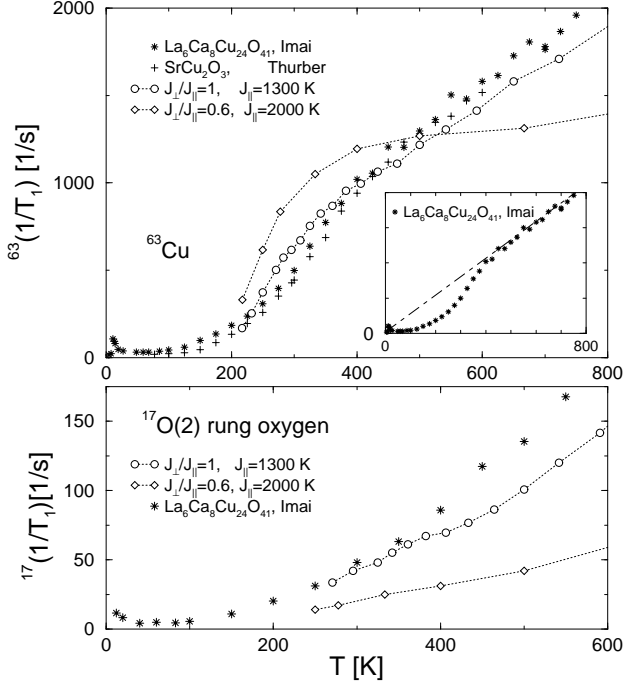}}
\caption[]{Comparison of NMR rates in ${\rm SrCu}_2{\rm O}_3$\cite{thurber}
and ${\rm La}_6{\rm Ca}_8{\rm Cu}_{24}{\rm O}_{41}$ \cite{imai} to numerical results.
Inset: Imai's data, the high-$T$ regime extrapolates to zero.}
\label{fig_exp}
\end{figure}
Let us now compare our results with experimental measurements of
$^{63}(1/T_1)$ and the rung oxygen $^{17}(1/T_1)$ rate
in ${\rm SrCu}_2{\rm O}_3$ and ${\rm La}_6{\rm Ca}_8{\rm Cu}_{24}{\rm O}_{41}$.
To convert our values to the experimental unit, we considered
the hyperfine couplings given in Ref.
\cite{ishida} for the $^{63}{\rm Cu}$ and in Ref. \cite{imai}
for the $^{17}{\rm O}$ nuclei, so that the remaining free parameters
are only $J_{\perp}$ and $J_{\|}$, which we have chosen according to those
obtained from the susceptibility fits in Fig. \ref{fig0}.
Considering that $T$ scales proportionally and $1/T_1$ inversely proportional
to $J_{\|}$, we found good overall magnitudes, however, the
precise $T$-dependence is not reproduced very accurately. 
Our results tend to indicate that $J_{\perp}/J_{\|}\sim 1$, in
disagreement with $J_{\perp}/J_{\|}\sim 0.5$ from the susceptibility.
Especially, the `linear' high-$T$ behavior of $^{63}(1/T_1)$ extrapolating to
zero (inset) cannot be reproduced with a smaller ratio $J_{\perp}/J_{\|}$.

These observations raise two important issues about ladder materials.
First, the hyperfine couplings measured in Refs.\cite{imai,ishida} may not be
sufficient for a quantitative discussion of NMR experiments within the
standard ladder model. A possible improvement may be to consider
transfered fields $^{63}F\propto(1+2r\cos q_x)^2+r'\cos^2(q_y/2)$ on the
copper atom, which can suppress the $(\pi,\pi)$-fluctuations by large
factors\cite{trans} and favor a smaller ratio $J_{\perp}/J_{\|}$.
This suggests that a better knowledge of the hyperfine couplings in ladder materials
is needed for a quantitative interpretation of measured NMR data.
Second, small corrections to the standard ladder such as interladder
or frustration couplings are known to have little effect on bulk properties
such as the spin gap\cite{wang_condmat}.
Therefore, the accessible susceptibility measurements ($T/\Delta\lesssim 1$) 
are relatively insensitive to such corrections. On the other hand,
local dynamic properties as NMR rates can be more substantially affected.

To summarize, we have shown that our technique provides reliable NMR rates
$1/T_1$ in the standard ladder over a wide temperature range. In particualar, we
demonstrated that the $q_y=\pi$ contributions are crucial for understanding
the crossover observed in $^{63}(1/T_1)$ experiments. Finally, we point out
how quantitative results can indicate inconsistencies in estimates of
$J_{\perp}/J_{\|}$ from suseptibility and NMR measurements.

We thank X. Zotos and M. Long for helpful suggestions, T. Imai for kindly
providing us the experimental measurements, D. C. Johnston and T. Xiang
for useful communications. Our work was supported by the Swiss National
Foundation grant no. 20-49486.96.


\begin{references}
\bibitem{defhyp} We have absorbed constants in the hyperfine couplings,
the link to Ref. \cite{imai} is given by:
$A^2=~^{63}\gamma^2_N/\mu^2_B\hbar(A^2_a+A^2_b)$, 
$G^2=~^{17}\gamma^2_N/\mu^2_B\hbar(G^2_a+G^2_b)$ when $G=C,D,F$.
\bibitem{omega} For all estimates of $1/T_1$, we consider
${\bar S}_{{\bar q}_x}(q_y,\omega\rightarrow 0)$.
\end{references}
\end{document}